\documentclass[10pt,twocolumn,superscriptaddress,showpacs,preprintnumbers,amsmath,amssymb,prb,aps]{revtex4}
\usepackage{graphicx}

\begin{document}


\title{Comparative study of the centrosymmetric and non-centrosymmetric superconducting phases of Re$_3$W using muon-spin-spectroscopy and heat capacity measurements}

\author{P. K. Biswas}
\email[]{P.K.Biswas@warwick.ac.uk}
\affiliation{Physics Department, University of Warwick, Coventry, CV4 7AL, United Kingdom}

\author{A. D. Hillier}
\affiliation{ISIS Facility, Science and Technology Facilities Council, Rutherford Appleton Laboratory, Chilton, Oxfordshire, OX11 0QX, U.K.}

\author{M. R. Lees}
\affiliation{Physics Department, University of Warwick, Coventry, CV4 7AL, United Kingdom}

\author{D. McK. Paul}
\affiliation{Physics Department, University of Warwick, Coventry, CV4 7AL, United Kingdom}

\date{\today}

\begin{abstract}
We compare the low-temperature electronic properties of the centrosymmetric (CS) and non-centrosymmetric (NCS) phases of Re$_3$W using muon-spin-spectroscopy and heat capacity measurements. The zero-field $\mu$SR results indicate that time reversal symmetry is preserved for both structures of Re$_3$W. Transverse-field muon spin rotation has been used to study the temperature dependence of the penetration depth $\lambda\left(T\right)$ in the mixed state. For both phases of Re$_3$W, $\lambda\left(T\right)$ can be explained using a single-gap $s$-wave BCS model. The magnetic penetration depth at zero temperature $\lambda\left(0\right)$ is $164(7)$ and $418(6)$~nm for the centrosymmetric and the non-centrosymmetric phases of Re$_3$W respectively. Low temperature specific heat data also provide evidence for an $s$-wave gap-symmetry for the two phases of Re$_3$W. Both the $\mu$SR and heat capacity data show that the CS material has a higher $T_c$ and a larger superconducting gap $\Delta(0)$ at 0~K than the NCS compound. The ratio $\Delta(0)/k_BT_c$ indicates that both phases of Re$_3$W should be considered as strong-coupling superconductors.  
\end{abstract}



\maketitle
\section{INTRODUCTION}
Since the discovery of the superconductivity in the heavy fermion CePt$_3$Si,~\cite{Bauer} non-centrosymmetric (NCS) superconductors have been the subject of intense theoretical and experimental investigation. The absence of a center of inversion symmetry in the crystal lattice means that antisymmetric spin-orbit coupling breaks parity. As a result, the superconducting pair wave function can have a mixture of spin-singlet and spin-triplet character.~\cite{Gorkov} This in turn may lead to unusual properties including a helical vortex phase,~\cite{Kaur} and complex phase diagrams involving superconductivity and magnetism.~\cite{Neupert} 

Novel physics has indeed been observed in many NCS superconductors. Examples include suppressed paramagnetic limiting or high upper critical fields~\cite{Frigeri,Mineev} in the heavy fermion materials CePt$_3$Si and CeRhSi$_3$,~\cite{Bauer,Kimura} and the transition metal compounds Nb$_{0.18}$Re$_{0.82}$ and Mo$_3$Al$_2$C,~\cite{Karki2, Karki} the appearance of superconductivity with antiferromagnetic order in CePt$_3$Si~\cite{Metoki} or at the border of ferromagnetism in UIr,~\cite{Akazawa} and time-reversal symmetry breaking in LaNiC$_2$.~\cite{Hillier} Superconductivity in non-centrosymmetric systems has also been reported in some  binary gallides containing iridium or rhodium,~\cite{Shibayama} under pressure in CeRhSi$_3$ and CeIrSi$_3$,~\cite{Kimura, Sugitani} and in the metal-rich borides Li$_2$(Pd$_{1-x}$Pt$_x$)$_3$B~\cite{Togano,Badica,Yuan} and Mg$_{10}$Ir$_{19}$B$_{16}$.~\cite{Klimczuk}

Superconductivity in Re$_3$W was first reported more then fifty years ago,~\cite{Blaugher1,Blaugher2} although since these initial reports there has been little published work on this material. This binary intermetallic compound contains heavy transition metals and so the spin-orbit coupling is expected to be strong. In addition, any complications arising from the presence of f-electron elements, such as strong electron correlations and the possibility of magnetically mediated superconductivity, are not expected. Recent powder neutron-diffraction studies have shown that Re$_3$W can adopt one of two different crystallographic structures.~\cite{Biswas} One phase has a NCS, cubic, $\alpha$-Mn structure,~\cite{Blaugher1,Blaugher2} is hard but brittle in nature, and has a superconducting transition temperature, $T_c$, of 7.8~K. The other phase is also superconducting but has a $T_c$ of 9.4~K, a hexagonal centrosymmetric (CS) structure, and is hard and malleable.  An as-cast CS sample can be converted to the NCS phase by annealing.  Remelting restores the CS structure. All this makes Re$_3$W a useful system in which to explore any differences in the superconducting state generated by switching from a CS to a NCS crystallographic structure.

In addition to $T_{c}$, we have already shown that many of the other superconducting properties of Re$_3$W change with the crystallographic structure.~\cite{Biswas} The value of the lower critical field, $H_{c1}$, is 97(1)~Oe for the NCS phase of Re$_3$W, while it is 279(11)~Oe for the CS phase. The temperature dependence of the upper critical field $H_{c2}(T)$ is different for the two materials, especially close to $T_{c}$. Well below $T_{c}$, giant flux jumps are observed in the magnetization versus applied magnetic field hysteresis loops of the CS phase, while in the NCS material no jumps are seen in the data and the magnetization becomes reversible in applied fields above $\sim10$~kOe.  

Some of these observations may be attributed to differences in the metallurgy of the CS and the NCS phases of Re$_3$W, while others are more likely to result from changes in the electronic properties, including variations in the magnitude and symmetry of the superconducting gap, as well as the temperature and field dependence of the gap below $T_{c}$. It is this latter possibility that we investigate in this work. Here, we report on a study of the superconducting properties of the NCS and the CS phases of Re$_3$W using muon spin relaxation/rotation ($\mu$SR). We also compare the $\mu$SR results with low-temperature heat capacity data collected on the same samples.

$\mu$SR is an ideal probe to study the superconducting state as it provides microscopic information on the local field distribution within the bulk of the sample. It can be used to detect small internal magnetic fields associated with the onset of an unconventional superconducting state~\cite{Aoki, Luke, Hillier} and to measure the temperature and field dependence of the London magnetic penetration depth, $\lambda$, in the vortex state of type-II superconductors.~\cite{Sonier, Brandt} The temperature and field dependence of $\lambda$ can in turn provide detailed information on the nature of the superconducting gap.

\section{EXPERIMENTAL DETAILS}

\subsection{SAMPLE PREPARATION}
Samples of the centrosymmetric phase of Re$_3$W were prepared by melting a stoichiometric mixture of Re lumps ($99.99\%$) and W pieces ($99.999\%$) in a high-purity Ar atmosphere on a water-cooled copper hearth using tungsten electrodes in an arc furnace. After the initial melt, the buttons were turned and remelted several times to ensure homogeneity. Samples of the non-centrosymmetric phase were made by annealing the as-cast samples for 5 days at $1500^\circ$C in a high-purity Ar atmosphere. Both samples contain small amounts of unidentified second phases.~\cite{Biswas} 

\subsection{$\mu$SR EXPERIMENTS}

Muon spin rotation ($\mu$SR) experiments were performed on the MuSR spectrometer of the ISIS pulsed muon facility, Rutherford Appleton Laboratory, UK. In the transverse field (TF) mode, an external magnetic field was applied perpendicular to the initial direction of the muon spin polarization. The magnetic field was applied above the superconducting transition and the sample then cooled to base temperature (FC). In this configuration the signals from the instrument's 64 detectors are reduced to two orthogonal components which are then fitted simultaneously. Data were also collected in zero field (ZF). Here, the decay positrons from the muons are detected and time stamped in the detectors which are positioned either before or after the sample. Using these counts, the asymmetry in the positron emission can be determined, and, therefore, the muon polarization is measured as a function of time.
In ZF mode, the stray fields at the sample position are canceled to within 1~mOe by three pairs of coils forming an active compensation system.

A powder sample of the NCS phase of Re$_3$W was mounted on a silver plate with a circular area of $\sim700$~mm$^2$. A small amount of General Electric (GE) varnish was added to the powdered sample in order to aid thermal contact. For the CS phase, several as-cast buttons were cut in half using a spark cutter. These hemispherical buttons were then partially remelted to form a circular disk with a cross sectional area of $\sim400$~mm$^2$. The disk was glued on to a silver plate using GE varnish. Thin silver foil was used to cover both samples. In the TF mode, any silver exposed to the muon beam gives a non-decaying sine wave. The sample and mount were then inserted into an Oxford Instruments He$^3$ sorption cryostat.

\subsection{HEAT CAPACITY EXPERIMENTS}

Heat capacity was measured using a two-tau relaxation method in a Quantum Design Physical Properties Measurement System at temperatures ranging from 1.9 to 300~K. The samples were attached to the stage using Apiezon N grease to ensure good thermal contact.

\section{EXPERIMENTAL RESULTS AND DISCUSSION}

\begin{figure}[tb!]
\begin{center}
\includegraphics[width=0.9\columnwidth]{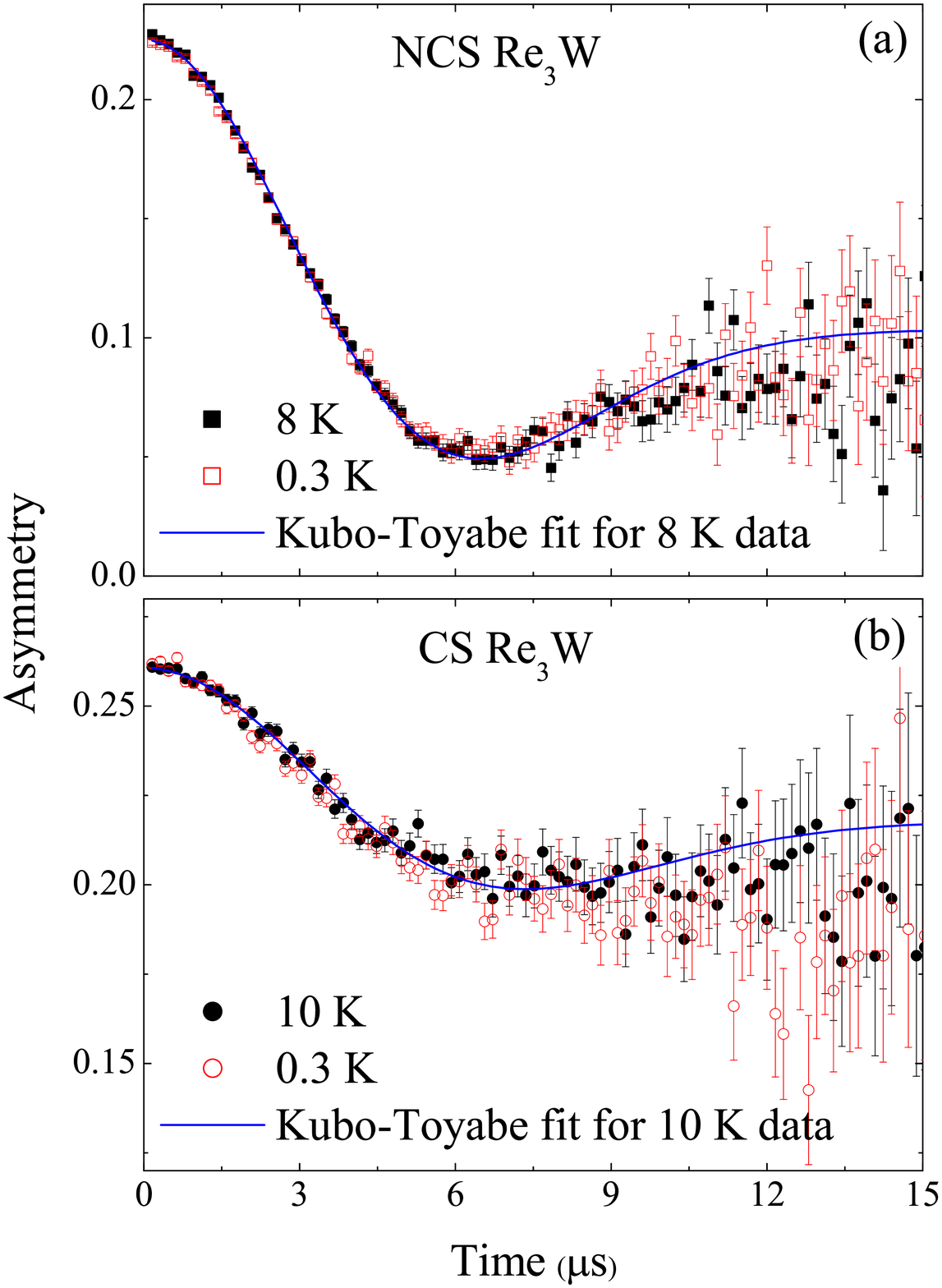}
\caption{\label{Figure1Biswas} (Color online) ZF-$\mu$SR time spectra collected at (a) 8 and 0.3~K for the NCS phase of Re$_3$W and (b) 10 and 0.3~K for the CS phase of Re$_3$W. The solid lines are fits to the data using the Kubo-Toyabe function as described in the text.}
\end{center}
\end{figure}

\begin{figure*}[tb!]
\begin{center}
\includegraphics[width=1.4\columnwidth]{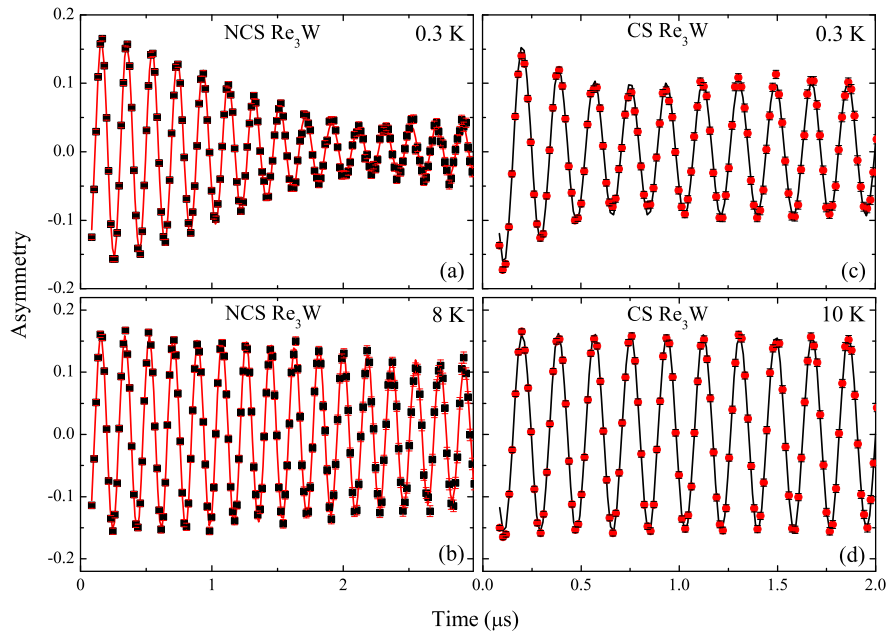}
\caption{\label{Figure2Biswas} (Color online) Transverse-field muon-time spectra (one component) collected in a magnetic field $H=400$~Oe at (a) 0.3 and (b) 8.0~K for the NCS phase of Re$_3$W and at (c) 0.3 and (d) 10~K for the CS phase of Re$_3$W.}
\end{center}
\end{figure*}

\begin{table}
\caption{Parameters extracted from the fits using the Kubo-Toyabe function to the zero-field-$\mu$SR data collected above and below $T_{c}$ for the non-centrosymmetric and centrosymmetric phases of Re$_3$W.}
\label{table_of_fits}
\begin{center}
\begin{tabular}[t]{lll}\hline\hline
{}~~~~~~~~&NCS Re$_3$W~~~~~~~~&CS Re$_3$W\\\hline
$A_{0}$ & 0.182$\pm$0.001 [8 K] & 0.064$\pm$0.001 [10 K]\\
{} & 0.178$\pm$0.001 [0.3 K]~~~~ & 0.069$\pm$0.002 [0.3 K]\\\hline
$\sigma$ ($\mu$s$^{-1}$) & 0.267$\pm$0.002 [8 K] & 0.235$\pm$0.004 [10 K]\\
{} & 0.266$\pm$0.002 [0.3 K] & 0.234$\pm$0.005 [0.3 K]\\\hline
$A_{bkgd}$ ~~~~~~~~& 0.043$\pm$0.001 [8 K] & 0.196$\pm$0.001 [10 K]\\
{} ~~~~~~~~& 0.047$\pm$0.001 [0.3 K] & 0.191$\pm$0.002 [0.3 K]\\\hline\hline
\end{tabular}
\par\medskip\footnotesize
\end{center}
\end{table}

We have performed a ZF-$\mu$SR study on both phases of Re$_3$W in order to search for any (weak) internal magnetism that may arise as a result of ordered magnetic moments, as well as to look for any temperature dependent relaxation processes associated with the onset of superconductivity.~\cite{Aoki, Luke, Hillier} Figures~\ref{Figure1Biswas}(a) and ~\ref{Figure1Biswas}(b) show the ZF-$\mu$SR signals from the NCS and CS phases of Re$_3$W respectively. There is no precessional signal and no obvious change in the observed relaxation rate between data collected above and below $T_{c}$ in either of the materials. The ZF data can be described by the Kubo-Toyabe function,~\cite{Hayano}

\begin{eqnarray}
\label{ZF_Fit}
G_z(t)=A_0\left[\frac{1}{3}+\frac{2}{3}(1-\sigma^{2}t^{2})\exp(-\frac{\sigma^{2}t^{2}}{2})\right]+A_{bkgd},
\end{eqnarray}

where $A_0$ is the initial asymmetry, $\sigma$ is the relaxation rate, and $A_{bkgd}$ is the background signal. The fits yield the parameters shown in Table~\ref{table_of_fits} with very similar values for each phase obtained above and below $T_{c}$. The observed behavior, and the values of $\sigma$ extracted from the fits, are commensurate with the presence of random local fields arising from the nuclear moments within the samples, that are static on the time scale of the muon precession.

There is no evidence for any spontaneous coherent internal fields at the muon sites arising from long-range magnetic order, in either the normal or the superconducting state. Nor are there any additional relaxation channels that may be associated with more exotic electronic phenomena such as the breaking of time-reversal symmetry.~\cite{Aoki, Luke, Hillier}  

\begin{figure}[tb!]
\begin{center}
\includegraphics[width=0.9\columnwidth]{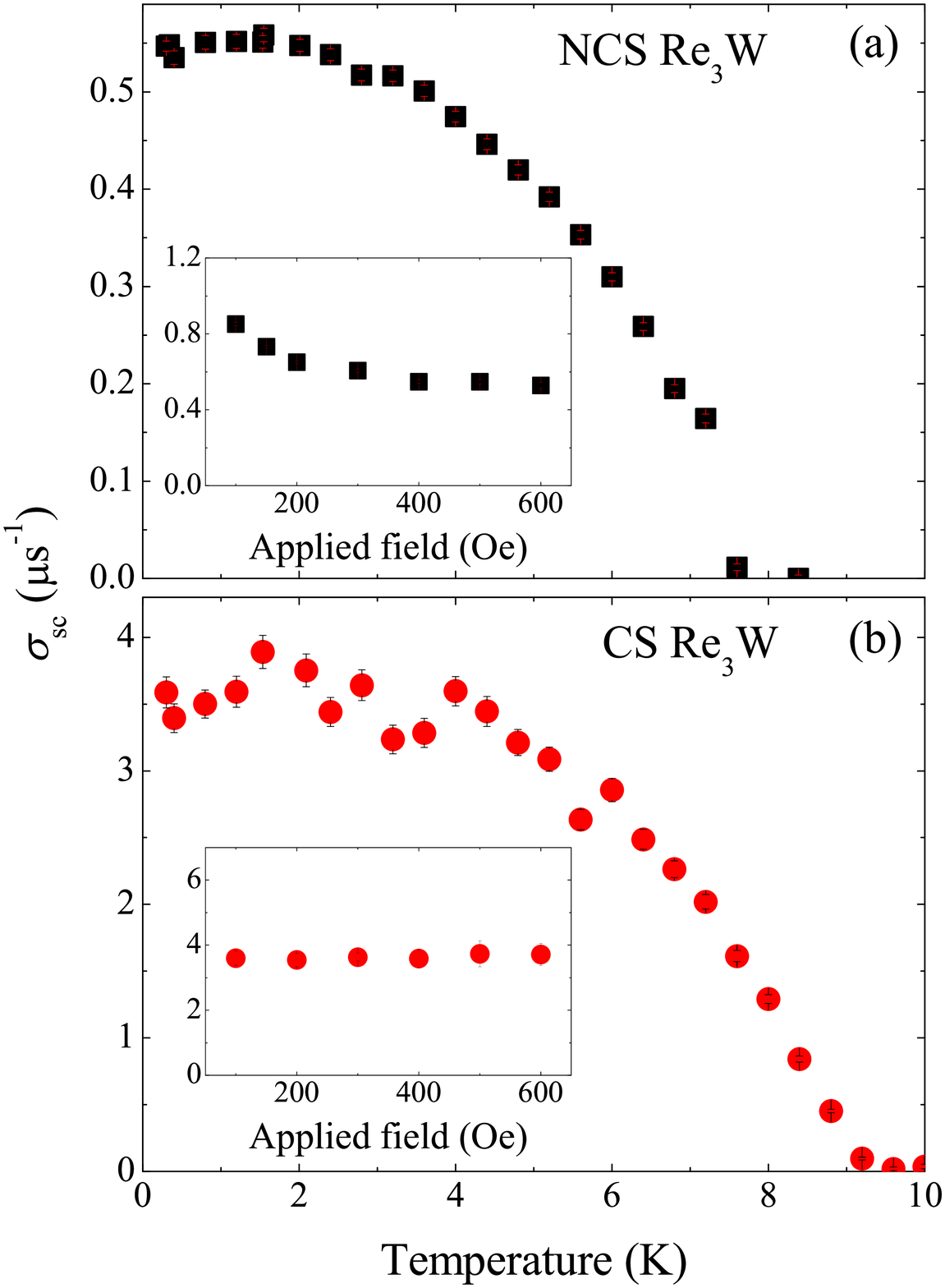}
\caption{\label{Figure3Biswas} (Color online) Temperature dependence of the superconducting muon spin relaxation rate $\sigma_{sc}$, collected in an applied magnetic field $H=400$~Oe for (a) the NCS and (b) the CS phase of Re$_3$W. The insets show the magnetic field dependence of $\sigma_{sc}$, obtained at 0.3~K for each of the phases.}
\end{center}
\end{figure}

Figure~\ref{Figure2Biswas} shows the TF-$\mu$SR precession signals below and above $T_c$ for both the NCS and the CS phases of Re$_3$W. The data were collected in an applied field of $H=400$~Oe to make sure that below $T_c$ the samples are in the mixed state. Figs.~\ref{Figure2Biswas}(b) and~\ref{Figure2Biswas}(d) show that in the normal state ($T>T_c$), the signals from both the phases of Re$_3$W decay very slowly, as there is a homogeneous magnetic field distribution throughout the samples. In contrast, Figs.~\ref{Figure2Biswas}(a) and~\ref{Figure2Biswas}(c) show that in the superconducting state ($T<T_c$), the decays are relatively fast due to the inhomogeneous field distribution from the flux-line lattice. The TF-$\mu$SR precession data were fitted using an oscillatory decaying Gaussian function,

\begin{eqnarray}
\label{Depolarization_Fit}
G_X(t)=A_{0}\exp\left(-\sigma^{2}t^{2}\right/2)\cos\left(\omega_{1} t +\phi\right)  \nonumber \\
+A_{1}\cos\left(\omega_{2} t +\phi\right),~
\end{eqnarray}
where $\omega_1$ and $\omega_2$ are the frequencies of the muon precession signal and background signal respectively, $\phi$ is the initial phase offset, and $\sigma$ is the Gaussian muon-spin relaxation rate. $\sigma$ can be written as $\sigma=\left(\sigma^{2}_{sc} + \sigma^{2}_{nm}\right)^{\frac{1}{2}}$, where $\sigma_{sc}$ is the superconducting contribution to the relaxation rate and $\sigma_{nm}$ is the nuclear magnetic dipolar contribution which is assumed to be constant over the temperature range of the study. Figs.~\ref{Figure3Biswas}(a) and ~\ref{Figure3Biswas}(b) show the temperature dependence of $\sigma_{sc}$ obtained in an applied TF of 400~Oe for the NCS and CS phases of Re$_3$W, respectively. The insets show the magnetic field dependence of $\sigma_{sc}$ at 0.3~K for each of the phases. $\sigma_{sc}$ is almost field independent for the CS phase, while there is an upturn in $\sigma_{sc}$ for the NCS phase of Re$_3$W at lower fields ($H<200$~Oe).~\cite{sigmanote}  These data confirm that in 400~Oe, $\sigma_{sc}$ is independent of the magnitude of the applied magnetic field.  

In a superconductor with a large upper critical field and a hexagonal Abrikosov vortex lattice, the Gaussian muon-spin depolarization rate $\sigma_{sc}$ is related to the penetration depth $\lambda$ by the expression

\begin{equation}
\frac{\sigma_{sc}^{2}\left(T\right)}{\gamma_\mu^2}=0.00371\frac{\Phi_0^2}{\lambda^{4}\left(T\right)},
\end{equation}

where $\gamma_\mu/2\pi=135.5$~MHz/T is the muon gyromagnetic ratio and $\Phi_0=2.068\times10^{-15}$~Wb is the flux quantum.\cite{Sonier,Brandt}

The magnetic penetration depths at $T=0$~K are found to be $\lambda_{NCS}\left(0\right)=418(6)$~nm for the NCS phase and $\lambda_{CS}\left(0\right)=164(7)$~nm for the CS phase of Re$_3$W.~\cite{Errors1} There is reasonable qualitative agreement between the penetration depths calculated from our $\mu$SR studies and those determined from dc magnetic susceptibility and rf tunnel diode resonator measurements, although the absolute values obtained from the $\mu$SR data are systematically higher than the $\lambda_{NCS}\left(0\right)$ of between 257(1) and 300(10)~nm reported for NCS Re$_3$W and the $\lambda_{CS}\left(0\right)=141(11)$~nm quoted for the CS material in earlier work.~\cite{Biswas,Zuev} 

The temperature dependence of the London magnetic penetration depth, $\lambda\left(T\right)$, can be calculated within the local London approximation~\cite{Tinkham,Prozorov} for an $s$-wave BCS superconductor in the clean limit using the following expression

\begin{equation}
\left[\frac{\lambda^{2}\left(0\right)}{\lambda^{2}\left(T\right)}\right]_{\rm{clean}}=1+2\int^{\infty}_{\Delta_{\left(T\right)}}\left(\frac{\partial f}{\partial E}\right)\frac{ EdE}{\sqrt{E^2-\Delta^2\left(T\right)}},
\end{equation}

where $f=\left[1+\exp\left(E/k_BT\right)\right]^{-1}$ is the Fermi function and $\Delta\left(T\right)=\Delta_{0}\delta\left(T/T_c\right)$. The temperature dependence of the gap is approximated by the expression~\cite{Fang,Errors1} $\delta\left(T/T_c\right)=\tanh\left\{1.82\left[1.018\left(T_c/T-1\right)\right]^{0.51}\right\}$ while in the dirty limit we have 
\begin{equation}
\left[\frac{\lambda^{2}\left(0\right)}{\lambda^{2}\left(T\right)}\right]_{\rm{dirty}}=\frac{\Delta\left(T\right)}{\Delta\left(0\right)}\tanh\left(\frac{\Delta\left(T\right)}{2k_BT}\right).
\end{equation}

\begin{figure}[tb!]
\begin{center}
\includegraphics[width=0.9\columnwidth]{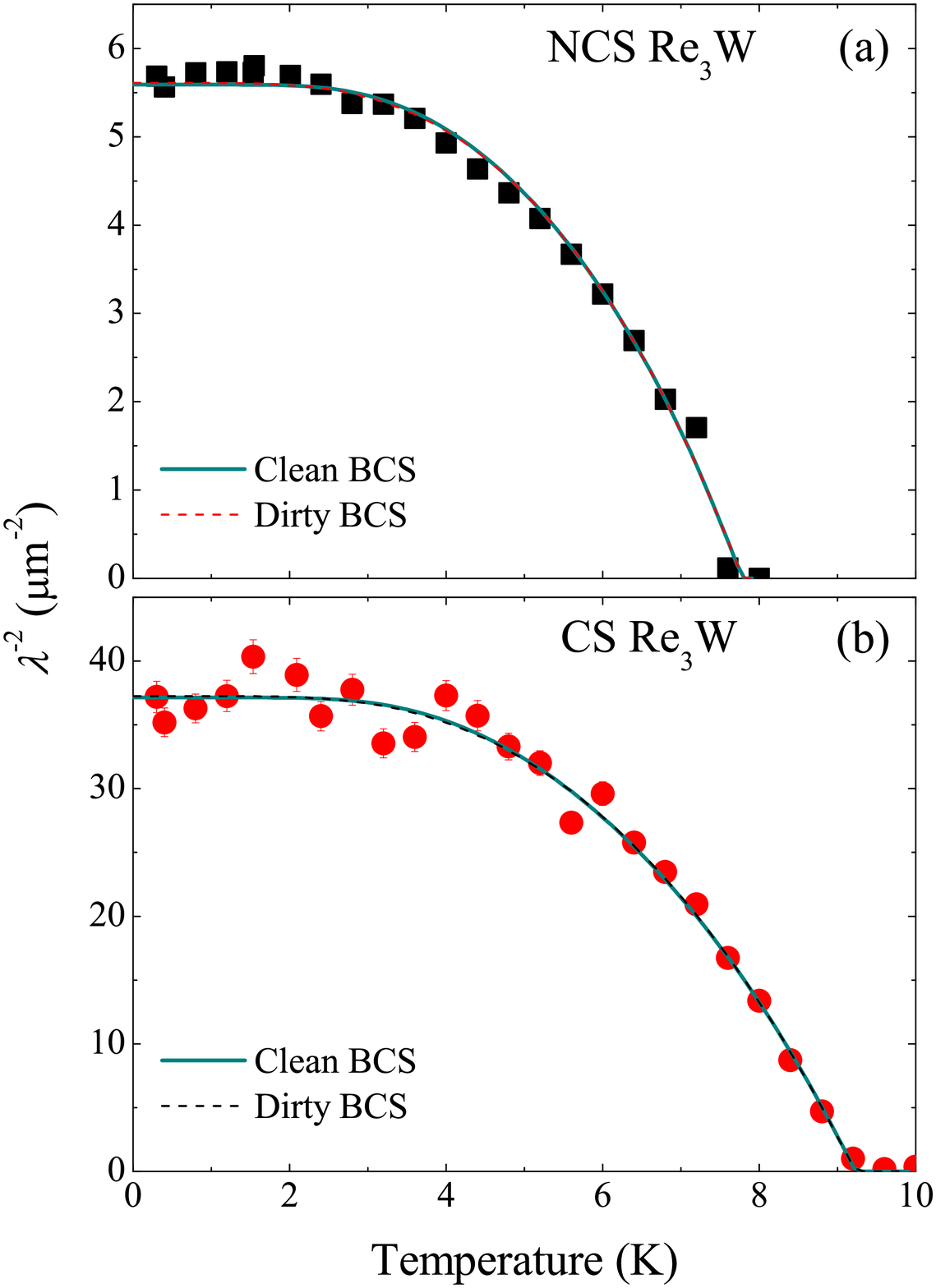}
\caption{\label{Figure4Biswas} (Color online) The inverse square of the London penetration depth $\lambda^{-2}$ as a function of temperature for (a) the NCS and (b) the CS phases of Re$_3$W. The solid lines are $s$-wave fits to the data.}
\end{center}
\end{figure}

We obtain good fits to the $\lambda^{-2}\left(T\right)$ data for the NCS and the CS phases using both the models discussed above (see Fig.~\ref{Figure4Biswas}). The parameters extracted from these fits are shown in Table~\ref{table_of_gapparameters}. There is little difference between the quality of the fits, as measured by $\chi^2_{norm}$, in the clean and dirty limits. As expected the magnitudes of the gap in the clean limit are larger than those obtained for the dirty limit but in both cases the values obtained place the materials in the strong-coupling limit.  

\begin{table}
\caption{Superconducting gap parameters extracted from the fits to the penetration depth data using a BCS model in the clean and the dirty limit for both the non-centrosymmetric and centrosymmetric phases of Re$_3$W.}
\label{table_of_gapparameters}
\begin{center}
\begin{tabular}[t]{llll}\hline\hline
NCS & Re$_3$W & & \\\hline
Model ~~& $\Delta$(0)~(meV)~~~& $\Delta$(0)/$k_{B}T_{c}$ ~~~& $\chi^2_{norm}$ \\\hline
Clean BCS ~~~& 1.49$\pm$0.04 & 2.22$\pm$0.06 & 1.74 \\
Dirty BCS ~~~& 1.38$\pm$0.07 & 2.05$\pm$0.10 & 1.72 \\ \hline\hline
CS & Re$_3$W & & \\\hline
Model ~~& $\Delta$(0)~(meV)~~~& $\Delta$(0)/$k_{B}T_{c}$ ~~~& $\chi^2_{norm}$ \\\hline
Clean BCS ~~~& 1.70$\pm$0.03 & 2.14$\pm$0.04 & 1.60 \\
Dirty BCS ~~~& 1.51$\pm$0.06 & 1.90$\pm$0.08 & 1.57 \\ \hline\hline
\end{tabular}
\par\medskip\footnotesize
\end{center}
\end{table}

\begin{figure}[tb!]
\begin{center}
\includegraphics[width=0.9\columnwidth]{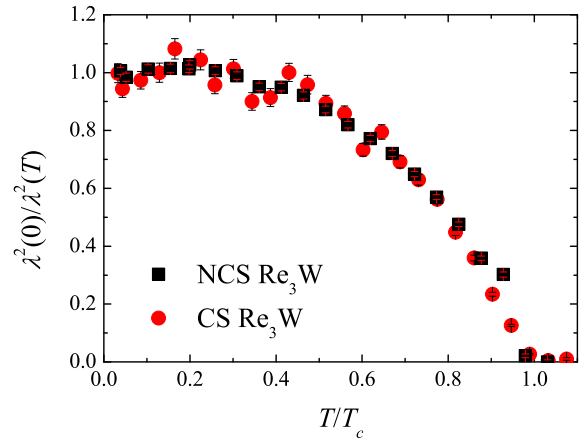}
\caption{\label{Figure5Biswas} (Color online) ${\lambda^{2}\left(0\right)}/{\lambda^{2}\left(T\right)}$ as a function of the reduced temperature $T/T_c$ for the CS and the NCS phases of Re$_3$W.}
\end{center}
\end{figure}

Figure~\ref{Figure5Biswas} shows ${\lambda^{2}\left(0\right)}/{\lambda^{2}\left(T\right)}$ as a function of the reduced temperature, $T/T_c$, for the CS and the NCS phases of Re$_3$W. The scaling of the data clearly suggests that both phases have the same gap symmetry.
Penetration depth measurements carried out on the NCS phase of Re$_3$W by the rf tunnel diode resonator technique and point-contact spectroscopy also suggest that the NCS phase of Re$_3$W is an $s$-wave superconductor, although Zuev \textit{et al}. could obtain good fits to their data for NCS Re$_3$W only in the dirty limit.~\cite{Zuev,Kuznetsova,Huang}

\begin{figure}[!htb]
\begin{center}
\includegraphics[width=0.9\columnwidth]{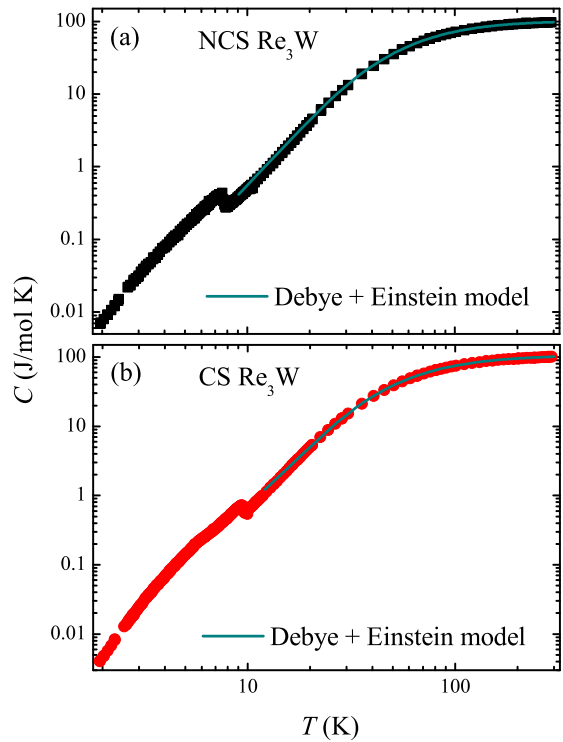}
\caption{\label{Figure6Biswas} (Color online) Temperature dependence of the specific heat $C$ for (a) the NCS phase and (b) the CS phase of Re$_3$W. Solid lines are the combined Debye - Einstein fits to the data.}
\end{center}
\end{figure}

To complement our $\mu$SR results we have carried out heat capacity measurements on the same samples used for the $\mu$SR experiments. Specific heat versus temperature $C(T)$ for both phases of Re$_3$W are shown in Fig.~\ref{Figure6Biswas}. As expected, no magnetic order could be detected down to 2 K and at high temperature, the signal is dominated by the contribution from the lattice. We can model the temperature dependence of these specific heat data using a single Debye term [see e.g. Ref.~\onlinecite{Gopal}]. For this analysis $\gamma_n$ is fixed to the value obtained from fits to the normal state data collected just above $T_c$ (see below). We obtain a Debye temperature $\Theta_D$ of 258(1) and 247(1)~K for the NCS and CS phases respectively. Better fits to these data can be achieved by adding an Einstein contribution to the total specific heat so

\begin{equation}
\label{specific_heat}
C(T)=\gamma_n T+ n_DC_D(T,\Theta_D)+n_EC_E(T,\Theta_E),
\end{equation} 
where $C_D$ and $C_E$ denote, respectively, the standard Debye and Einstein contributions to the specific heat with weighting fractions of $n_D$ and $n_E$.~\cite{Gopal} These fits produce the values for $\Theta_D$ and $\Theta_E$ shown in Table~\ref{table_specific_heat}.

\begin{figure}[!htb]
\begin{center}
\includegraphics[width=0.9\columnwidth]{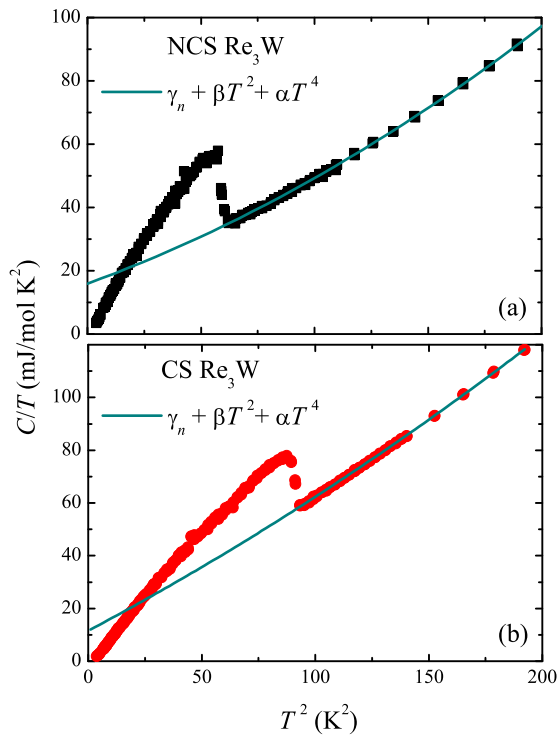}
\caption{\label{Figure7Biswas} (Color online) $C$/$T$ versus $T^2$ for (a) the NCS phase and (b) the CS phase of Re$_3$W. Solid lines are fits to the low-temperature data above $T_c$ using Eq.~\ref{lowHC}.}
\end{center}
\end{figure}

Figure~\ref{Figure7Biswas} shows $C/T$ versus $T^2$ for the two phases of Re$_3$W. Jumps in the specific heat data due to superconducting phase transitions are clearly observed at 7.8 and 9.4~K for the NCS and CS phases of Re$_3$W. These values are in good agreement with both the $\mu$SR data and our previously published magnetization and transport data.~\cite{Biswas} The data once again confirm the bulk nature of the superconductivity. 

In principle it is possible to obtain the lattice contribution to the specific heat in the superconducting state by applying a large enough magnetic field to drive the sample into normal state. However, the upper critical fields of both phases of Re$_3$W are high, making it impossible for us to access the normal state at low temperatures.~\cite{Biswas} Nevertheless, a fair estimate of the lattice contribution can be made by fitting the low-temperature specific heat data in the normal state just above $T_c$ using   

\begin{equation}
\label{lowHC}
C(T)=\gamma_n{T}+\beta{T}^3+\alpha{T}^5,
\end{equation} 
where $\gamma_n{T}$ is the electronic contribution and $\beta{T}^3+\alpha{T}^5$ represents the phonon contribution to the specific heat.
The solid lines in Figure~\ref{Figure7Biswas} show the fits to the normal state data using equation~\ref{lowHC}. The fit parameters are listed in Table~\ref{table_specific_heat}. The $\gamma_n$ values we obtain are consistent with a previous heat capacity study of a sample of Re$_3$W that was a mixture of the NCS and CS materials.~\cite{Jing} 

\begin{figure}[!htb]
\begin{center}
\includegraphics[width=0.9\columnwidth]{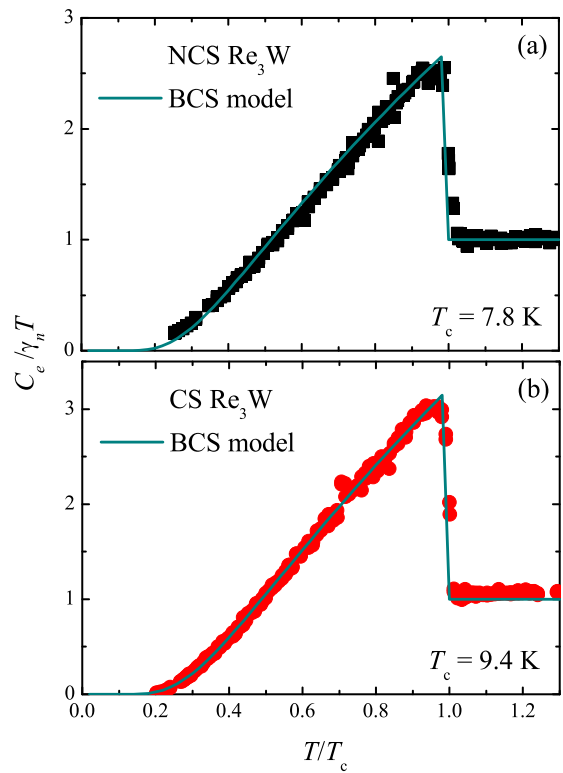}
\caption{\label{Figure8Biswas} (Color online) Electronic contribution to the specific heat $C_e$/$\gamma_n{T}$ plotted as a function $T$/$T_c$ for (a) the NCS phase and (b) the CS phase of Re$_3$W. Solid lines are fits to the data using a single-gap BCS model.}
\end{center}
\end{figure}

The electronic contribution to the specific heat $C_e$ can then be obtained by subtracting the phonon contribution from the total specific heat data. Figure~\ref{Figure8Biswas} shows the normalized electronic specific heat, $C_e$/$\gamma_n{T}$ of the NCS and CS phases of Re$_3$W as a function of reduced temperature $T$/$T_c$. To perform a fit to the $C_e/\gamma_n{T}$ data in the superconducting state, we use the single-gap BCS expression~\cite{Padamsee} for the normalized entropy $S$,
\begin{equation}
\label{two_gap_hc1}
\frac{S}{\gamma_{n}{T_c}}=-\frac{6}{\pi^2}\frac{\Delta(0)}{k_{B}T_{c}}\int^{\infty}_{0}[f\ln{f}+(1-f)\ln(1-f)]dy,
\end{equation}
with $f=\left[\exp\left(E/k_BT)\right)+1\right]^{-1}$, $E=\left[\epsilon^2+\Delta^2(t)\right]$ where $\epsilon$ is the energy of the normal electrons measured relative to the Fermi energy, $y=\epsilon/\Delta(0)$, and $t=T/T_c$ is the reduced temperature.
The specific heat $C$ is then given by
\begin{equation}
\label{two_gap_hc2}
\frac{C}{\gamma_{n}T_{c}}=t\frac{d(S/\gamma_{n}T_{c})}{dt}.
\end{equation}
The temperature dependence of the energy gap varies as $\Delta(t)=\Delta(0)\delta(t)$, where $\delta(t)$ is the temperature dependence of the normalized BCS gap.~\cite{Muhlschlegel} The solid lines in Fig.~\ref{Figure8Biswas} are the fits to the data using this single-gap BCS model. From these fits we obtain the superconducting gap parameters listed in Table~\ref{table_specific_heat}. There is good agreement between the gap parameters obtained from the $\mu$SR and the heat capacity measurements. These result confirm that both phases of Re$_3$W should be considered as strong-coupling superconductors.

\begin{table}
\caption{Parameters for the non-centrosymmetric and centrosymmetric phases of Re$_3$W obtain from fits to the high and low temperature specific heat data (see text for details).}
\label{table_specific_heat}
\begin{center}
\begin{tabular}[t]{lll}\hline\hline
{}~~~~~~~~&NCS Re$_3$W~~~~~~~~&CS Re$_3$W\\\hline
$\Theta_{D}$ (K) $[n_D]$ & 228$\pm$6 [0.78] & 219$\pm$1 [0.81]\\
$\Theta_{E}$ (K) $[n_E]$~~~~~~~~& 292$\pm$15 [0.22] & 333$\pm$6 [0.19]\\
$\gamma_n$ (mJ/mol K$^2$) & 15.9$\pm$0.6 & 11.6$\pm$0.8\\
$\beta$ (mJ/mol K$^4$) & 0.26$\pm$0.01 & 0.45$\pm$0.01\\
$\alpha$ ($\mu$J/mol K$^6$) & $0.73\pm0.04$ & $0.51\pm0.04$\\
$\Delta(0)$ (meV) & 1.25$\pm$0.01 & 1.56$\pm$0.01\\
$\Delta(0)/k_{B}T_c$ & 1.85$\pm$0.02 & 1.90$\pm$0.02\\\hline\hline
\end{tabular}
\par\medskip\footnotesize
\end{center}
\end{table}

\section{CONCLUSIONS}
We have performed $\mu$SR and specific heat studies on polycrystalline samples of both the centrosymmetric and the non-centrosymmetric superconducting phases of Re$_3$W. There is no evidence in either material for any long-range magnetic order, nor for any unusual electronic behavior arising from the non-centrosymmetric structure. Our results confirm that time-reversal symmetry is preserved in this system. The absolute values of the magnetic penetration depth are $\lambda_{NCS}\left(0\right)=418(6)$~nm and $\lambda_{CS}\left(0\right)=164(7)$~nm. Interestingly, the change in structure appears to have no effect on either the symmetry or the temperature dependence of the superconducting gap. The temperature dependence of $\lambda$ for both structural phases of Re$_3$W can be described using a single gap $s$-wave BCS model. The $s$-wave symmetry present in the NCS phase of Re$_3$W is not unprecedented. A conventional s-wave behavior has also been observed in several other NCS superconductors.~\cite{Togano,Badica,Klimczuk,Shibayama,Karki,Karki2} Heat capacity measurements on the same samples confirm the $\mu$SR results. The magnitudes of the superconducting gaps obtained from both the $\mu$SR and the specific heat studies suggest that both materials are strong-coupling superconductors.

\begin{acknowledgments}
PKB would like to thank the Midlands Physics Alliance Graduate School (MPAGS) for sponsorship. Some of the equipment used in this research was obtained through the Science City Advanced Materials project: Creating and Characterizing Next Generation Advanced Materials project, with support from Advantage West Midlands (AWM) and part funded by the European Regional Development Fund (ERDF).     
\end{acknowledgments}

\bibliography{Biswas}

\begin{thebibliography}{38}
\bibitem[{sig()}]{sigmanote}
\bibinfo{note}{$H_{c1}$ is $\sim97$~Oe for NCS Re$_3$W and $\sim279$~Oe for CS
  Re$_3$W. So the lower values of $H$ shown in the insets of
  Figure~\ref{Figure3Biswas} are $\leq H_{c1}$. The flux pinning in the CS
  material is much stronger than in the NCS phase.~\cite{Biswas} Therefore, the
  most likely cause of the upturn in $\sigma_{sc}$ in low fields for the NCS
  phase is flux exclusion.}
\bibitem[{Err()}]{Errors1}
\bibinfo{note}{The error in $\lambda(0)$ is the statistical error arising from
  the fit to the $\lambda^{-2}(T)$ data using the model described in the text.
  The error quoted does not take into account any systematic errors (e.g.
  vortex lattice disorder) that may be present in the data.}
\end{thebibliography}

\end{document}